\documentclass[aps,prl,superscriptaddress,showpacs]{revtex4}
\usepackage[dvips]{graphicx}
\usepackage{mathptmx}
\usepackage{verbatim}

\begin{document}
\title{Electronic Highways in Bilayer Graphene}
\author{Zhenhua Qiao$^\dag$}
\affiliation{Department of Physics, The University of Texas at
Austin, Austin, Texas 78712, USA}
\author{Jeil Jung$^\ddag$}
\affiliation{Department of Physics, The University of Texas at
Austin, Austin, Texas 78712, USA}
\author{Qian Niu$^\S$}
\affiliation{Department of Physics, The University of Texas at
Austin, Austin, Texas 78712, USA}\affiliation{International Center for Quantum Materials, Peking University, Beijing 100871, China}
\author{Allan H. MacDonald}
\affiliation{Department of Physics, The University of Texas at
Austin, Austin, Texas 78712, USA}
\date{\today{}}

\begin{abstract}
Bilayer graphene with an interlayer potential difference has an
energy gap and, when the potential difference varies spatially,
topologically protected one-dimensional states localized along the
difference's zero-lines. When disorder is absent, electronic travel
directions along zero-line trajectories are fixed by valley Hall
properties. Using the Landauer-B\"uttiker formula and the
non-equilibrium Green's function technique we demonstrate
numerically that collisions between electrons traveling in opposite
directions, due to either disorder or changes in path direction, are
strongly suppressed. We find that extremely long mean free paths of
the order of hundreds of microns can be expected in relatively clean
samples. This finding suggests the possibility of designing low
power nanoscale electronic devices in which transport paths are
controlled by gates which alter the inter-layer potential landscape.
\end{abstract}

\maketitle

\begin{figure}
\includegraphics[width=5cm,angle=270]{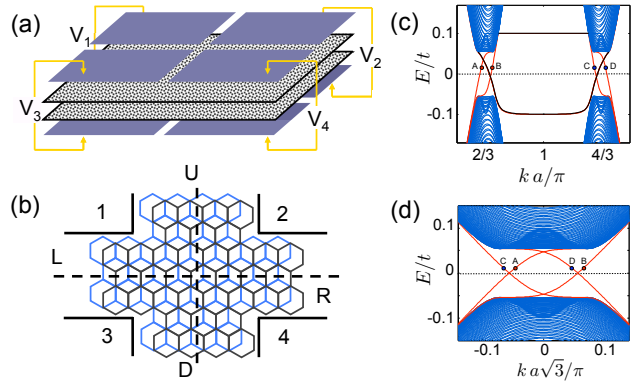}
\caption{(Color Online) (a) Model device with four regions that can
be gated to positive or negative inter-layer potential values. (b)
Schematic representation of the lattice geometry used in our
numerical simulations. The horizontal and vertical axes are chosen
to be aligned along the zigzag and armchair honeycomb lattice
directions of both the device and the graphene bilayer reservoirs
used in our four-probe NEGF calculations. (c)  One-dimensional band
structure of a zigzag ribbon in which the inter-layer potentials of
Eq.~1 change sign at the ribbon center: $U_{i} = \pm V(x) = \pm 0.1
\, t \, {\rm sgn}(x)$. Two 1D modes traveling in each direction are
spatially localized at the ribbon center. Right going states are
labeled with letters A, B whereas left going states are labeled as
C, D. Additional 1D channels appear in the gap that are localized
near opposite edges of the ribbon. These states are plotted in black
and are doubly degenerate due to inversion symmetry across the
ribbon. (d) The band structure in the armchair case has two pairs of
oppositely propagating channels located at the ribbon center but
does not support edge states. The atomically scale sharp potential
variation we used leads to a small avoided crossing gap $\Delta \sim
0.0014 t$ at the neutrality point.  The size of this gap shrinks
rapidly when the potential variation becomes smoother. }
\label{Setup}
\end{figure}

More than half a decade after seminal transport studies of graphene
sheets demonstrated the material's half-quantized quantum Hall
effect~\cite{novoselov,yuanbo}, interest is turning toward
applications of the material's exceptional
properties~\cite{geim,castroneto}. The half-quantized Hall effect in
graphene is a manifestation of momentum-space Berry phases
associated with its sublattice pseudospin
degree-of-freedom~\cite{beenaker,beenaker2,pseudospin1,pseudospin2,pseudospin3}.
The physics explored in the present paper centers on one dimensional
(1D) states which have the same origin as those that appear in
bilayer graphene~\cite{morpurgo}, monolayer graphene~\cite{yao}, and
chirally stacked multilayer graphene~\cite{xiaofan,chiral} along
zero-lines of inversion-symmetry-breaking potentials. When uniform,
these potentials open gaps and induce quantized but canceling
quantized Hall responses from $\pi$-orbital electronic states near
$K$ and $K'$ valleys. The quantized Hall response can be calculated
by integrating the momentum-space Berry curvatures over occupied
valence band states~\cite{morpurgo, yao, xiaofan}, and changes sign
when the inversion-symmetry-breaking potential changes sign. The 1D
states we study are closely related to the edge states which are
always present at spatial boundaries between regions with different
quantized Hall conductances. Because the valley Hall conductivity in
chirally stacked $N$ layer graphene, $\sigma^v_{xy} = N e^2/2h$, the
number of interface channels per valley localized along a zero-line
is also equal to $N$~\cite{chiral}. Similarly, because opposite
valleys have opposite Hall conductance sign, the chiral edge states
associated with different valleys travel in different directions.
Below we refer to the 1D states localized near the zero-lines as
{\em kink} states.

The possibility of controlling current paths in graphene with gates
has been explored previously by studying bipolar p-n junctions and
unipolar fibre-optic guides~\cite{marcus}. The present work shows
that kink state conducting channels, which can be formed simply by
gating bilayer graphene, are nearly ballistic in clean samples and
have very long mean free paths when the disorder strength is small
compared to the 2D bulk band gap. In addition, the kink states
retain substantial valley pseudospin memory at bends and at
intersections of zero-line trajectories. Our calculations strongly
suggest that pseudospin electronics \cite{beenaker} can be realized
by controlling internal zero-line trajectories.
As long as a kink-state electron maintains its valley label, its
direction of travel along a zero-line is uniquely determined.
Transport properties are then completely determined by zero-line
topology. In the case of $N=2$ bilayer graphene the symmetry
breaking potential is simply the electric potential difference
between layers~\cite{mccann}, which is easily altered by gates as
illustrated in \ref{Setup}(a).
This setup constitutes the simplest example of the valley valve proposal
consisting of two valley filters placed sequentially.  \cite{morpurgo}

Since valley label is a good quantum number only in the absence of
disorder, and then only for straight zero-lines that do not follow
armchair directions, it is critical to address the robustness of
valley memory. In this article we report on a numerical
non-equilibrium Green's function (NEGF) study of the influence of
bends and disorder on kink state transport properties. We employ a
$\pi$-band tight-binding model rather than a continuum
model~\cite{morpurgo, arun} which automatically conserves valley
index. We find that current paths nevertheless follow continuum
model predictions to a remarkable degree. This property still holds
when portions of the zero-line follow one of the armchair directions
even though armchair ribbons do not support edge states at the
ribbon/vacuum boundary\cite{jian_nature,chiral}.

The band structures of straight zigzag and armchair ribbons with a
zero-line along the ribbon center are illustrated in panel (c) and
(d) of \ref{Setup}. The anticipated pair of kink states localized at
the sample center appears for both ribbon orientations. For the
zigzag case in panel (c), kink states appear near ribbon wavevectors
$k= 2 \pi/3a, 4 \pi/3a$ where $a=2.46 {\rm \AA}$ is the lattice
constant of graphene, as suggested by a bulk graphene band
projection\cite{chiral}. Zigzag edges support edge state channels
localized at the ribbon/vacuum edges in addition to the kink states,
whereas armchairs ribbons do not support edge states and all kink
states appear near 1D momentum $k=0$. The close proximity of
opposite-velocity kink states in both real-space and 1D momentum
space might suggest that the continuum model picture should fail
badly when the zero-line direction is close to an armchair
direction. We will show that this is {\em not} the case.


\begin{figure*}
\includegraphics[width=16cm]{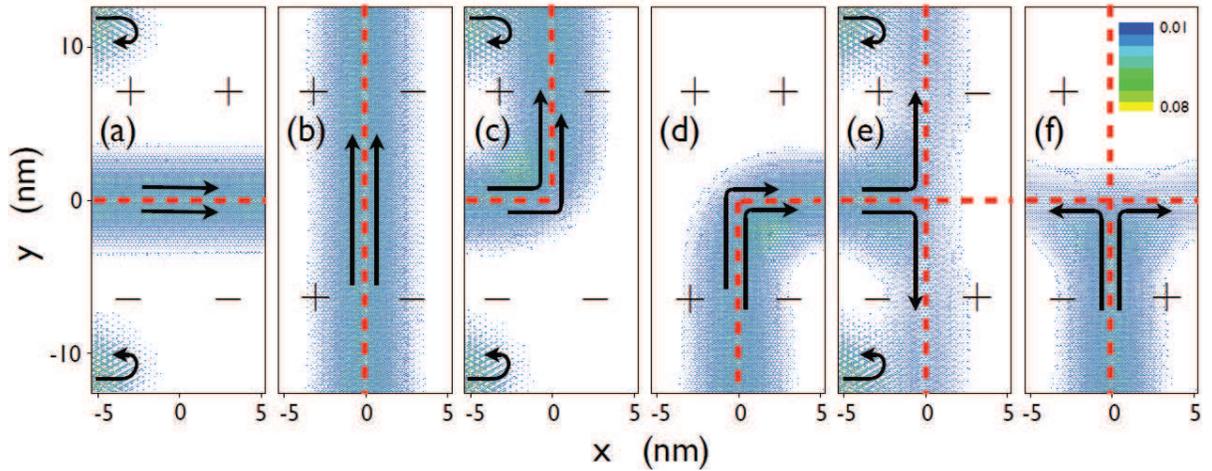}
\caption{(Color Online) Fermi-level kink scattering-state density
distributions in four-probe bilayer graphene cross-bars for
representative potential difference zero-line configurations. The
left, right, up, and down reservoirs each have a single zero-line at
the ribbon center which matches a system zero line at its boundary.
Each point in these plots corresponds to an individual carbon atom
site. The crystal is oriented so that the $x$ axis is along a zigzag
direction and the $y$ axis along an armchair direction. The thick
dotted lines indicate the potential difference zero-lines while the
$\pm$ symbols specify the potential difference sign in four
quadrants imagined to be controlled by four separate front/back gate
pairs. The panels (a), (c), and (d) are for scattering states
incident from the left lead while (b), (d), and (f) are for
scattering states incident from the lower lead. Transmission along
an isolated zero [(a-d)] line is essentially perfect.  When two
zero-lines intersect [(e-f)], the current splits into contributions
that follow the two available outgoing paths. In all cases chirality
is preserved, even when the path bends.  In cases (a-e) the
low-potential region is always on the right in the direction of
travel while in case (f) it is always on the left. These properties
are satisfied equally well for zigzag and armchair kinks. }
\label{currents}
\end{figure*}

Kink states have definite chirality if they preserve their valley
labels; states in one valley propagate along zero-lines keeping
low-potential regions on the left, while states in the other valley
keep low-potential regions on the right. Zero-lines intersect when
the potential difference landscape has a zero saddle point. For a
general continuous potential-difference profile, a system can have
many zero lines some of which are closed. When valley memory is
perfectly retained, only open paths connected to reservoir
zero-lines are relevant for transport. Zero-line considerations are
therefore relevant to the analysis of transport in neutral systems
with smooth random potential differences. In this paper, however, we
concentrate on systems with simple gate-defined potential-difference
profiles designed to control current paths in bilayer graphene
systems.

We model the case in which the leads are bilayer graphene ribbons
with a single zero-line at their centers, and the system is divided
into four quadrants in which the sign of the potential difference
can be varied independently as indicated schematically in
\ref{Setup}(a). Both incoming and outgoing states in the leads
therefore have definite pseudospin labels. If pseudospin memory were
perfect, injected electrons would travel following the continuation
of the lead's zero-line to one of the reservoirs.

Our explicit calculations are based on the $\pi$-orbital
tight-binding model:
\begin{eqnarray}
H =   -\sum_{\langle i,j \rangle}  \gamma_{i,j} \,\,  c_i^{\dag} c_j
+ \sum_{i} U_{i} \, \, c_{i}^{\dag} c_{i}, \label{eq1}
\end{eqnarray}
where $U_{i}$ is a $\pi$-orbital site energy and $\gamma_{i,j}$ is
either a nearest-neighbor in-plane hopping amplitude with value $ t
= 2.6 \,\, eV$ or a vertical inter-layer hopping amplitude with
value $t_{\perp} = 0.34 \,\, eV$.
The trigonal warping $\gamma_3 \sim 0.1 t$ term can play a role in the limit of vanishingly small gaps
\cite{alvaro} but are unimportant for our present discussion, as we show in the supporting information.
Here $c^{\dag}_i$ and $c_{i}$ are
$\pi$-orbital creation and annihilation operators for site $i$. In
most of our numerical simulations, we have considered a bilayer
graphene flake containing a total of 120 $\times$ 92$\times$2 =
22080  (vertical, horizontal, layer) atomic sites in the central
scattering region, corresponding to a few hundreds of ${\rm n m}^2$
of  flake area. We create kink states by setting $U_i \to \pm 0.1 \,
t$  so that the sum of site energies in different layers is
everywhere zero and the difference is $\pm 0.2 t$ in the $\pm$
regions.  The atomic scale variation of the potential difference is
not physically realistic, of course, since the sharpness of its
spatial profile cannot exceed the greater of the physical gate
separation and the vertical distance between bilayer and gate. In
the supplementary information we show that our results are not
altered in any essential way as long as the distance over which the
potential difference shifts between positive and negative values is
smaller than $\sim 100$ nm. The potential differences open up gaps
in the spectrum so that the
only states   
at the Fermi level of a nearly neutral bilayer are ribbon edge
states and kink states localized along zero-lines. We label the four
semi-infinite bilayer graphene reservoirs in our NEGF calculations
up (U), down (D), left (L) and right (R).

Our main numerical results, summarized in \ref{currents}, were
obtained for model flakes with zigzag edges in the horizontal
direction and armchair edges in the vertical direction. By varying
the gating potentials we can arrange to have vertical or horizontal
zero lines in the system, to have a single zero line that rotates by
$90^{\circ}$ between zigzag and armchair directions, or to have two
such zero lines that intersect at the middle of the sample.
Configurations like this, in which the zero lines of interest do not
intersect with the edge of the system can be used to isolate kink
state conducting channels from edge state conducting channels. We
anticipate that disorder at the edges will tend to localize edge
state transport. For the devices that we have in mind, increasing
disorder at the edges may in fact be desirable in order to mitigate
their possible role in transport.

We study how controlling the potential-difference profile can
control transport properties by calculating the conductances between
probes for each gating geometry. The conductance $G_{pq}$ from the
$q$-th probe to the $p$-th probe can be evaluated from the
Landauer-B\"{u}ttiker formula:
\begin{equation}
{\rm G}_{pq}=({e^2}/{h}) \; {\rm {Tr}}[\Gamma _{p} {\rm G}^r \Gamma
_{q} {\rm G}^a],
\end{equation}
where ${\rm G}^{r/a}$ are the retarded and the advanced Green
functions of the central scattering regime which we evaluate
following the same procedure as outlined in earlier
work~\cite{zhenhua}, with a small shift of the Fermi energy from
neutrality ($E=0.001 \, t$) to prevent the small avoided crossing
gaps in the armchair direction from playing a role. All quantities
are matrices with system carbon site labels. Here $\Gamma _p$ is the
line-width function coupling the $p$-th probe to the scattering
region, and can be calculated from the self-energy of the
semi-infinite lead using the transfer matrix
method~\cite{mtransfer}. Our results are summarized visually in
\ref{currents} by plotting the local density of states contribution
$\rho_p({\bf r},\varepsilon_F)$  of  scatttering states injected
from $p$-th probe. These are calculated using the
formula~\cite{datta}:
\begin{equation}
\rho_{p}({\bf r},\varepsilon_F)=\frac{1}{2\pi}[{\rm G}^r \Gamma _{p}
{\rm G}^a]_{{\bf r} {\bf r}}.
\end{equation}

We consider first the simplest setup, which has an isolated
horizontal zero-line. Incoming electrons propagate ballistically
across the system. The horizontal conductance is almost exactly two
in units of $e^2/h$ as expected for this disorder free, straight
zero-line case. For the horizontal zigzag orientation case
illustrated in \ref{currents}(a), we see that a pair of edge
channels appears in addition to the kink states, as
expected~\cite{castro,jian_nature,chiral}, for zigzag edge
boundaries. Electrons incident upon the system in these edge channel
states are reflected when they encounter the system's impenetrable
gapped region. In the case of a straight vertical zero-line in
\ref{currents}(b), current flows only through the kink channels
because there are no edge states associated with the armchair edge
terminations.

\begin{figure*}
\includegraphics[width=16cm]{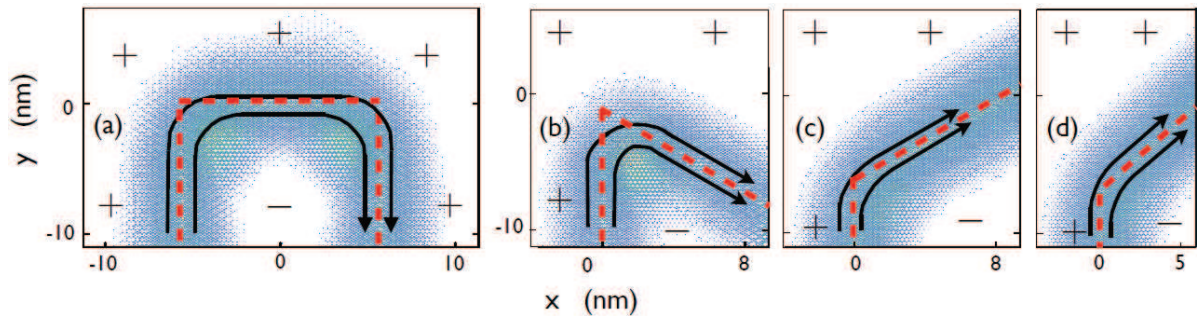}
\caption{ (Color Online) A few examples of transport along zero
lines with a variety of turn angles. In case (a) the zero line
currents are reversed by consecutive  $90^{\circ}$ turns changing
the propagation direction first from armchair to zigzag and then
from zigzag to armchair. We also show numerical results for turn
angles of $60^{\circ}$ (b), and $120^{\circ}$ (c)  from armchair to
armchair directions. In case (d) the turn angle is $131^{\circ}$ and
the final propagation path is neither zigzag nor armchair. The
numerically calculated conductances between the contacts indicate
almost perfect transmission along the zero-lines with no
backscattering due to the finite turn angles. } \label{moreturns}
\end{figure*}

The second setup  [panels (c) and (d)] supports an isolated
zero-line which bends by $90^{\circ}$. Pseudospin index selective
current propagation is demonstrated to be even more robust in this
device geometry in which three system quadrants have the same
difference-potential. The scattering state charge density profiles
and calculated conductances show that an incoming kink current
approaching the bend point makes a ninety degree turn in order to
preserve its pseudospin, instead of changing its valley index and
reflecting. For currents coming from the $L$ probe and making an
upward turn, we obtained transmission conductances of $G_{U/R/D,L} =
(1.97,0.01,0.00) $, whereas for an incoming current from the $D$
contact making a right turn the calculated conductances are
$G_{L/U/R,D} = (0.01, 0.00, 1.97)$. The small deviation of our
numerical results from the ideal transmission of two units of
$e^2/h$ is attributed in the absence of disorder to the finite size
of our simulation cell that cannot capture the entirety of the kink
state wave function tails away from the domain interface. For these
$90^{\circ}$ turns the kink state propagation direction changes
either from zigzag to armchair or from armchair to zigag.
Nevertheless there is almost no current leaking towards the other
leads in the device because there are no zero-line conduction
panels.

The third setup examines the alternating potential-difference
profile case  [see panels (e) and (f)] in which two zero-lines
intersect. Our calculations show that this arrangement creates an
electronic beam splitter. Propagation in the forward direction at
the intersection point is forbidden by the pseudospin filtering rule
since that direction of travel requires that the pseudospin be
reversed. Our numerical calculations show conductances of
$G_{U/R/D,L} = (0.99, 0.02, 0.99)$ for currents entering through the
probe $L$ and $G_{L/U/R,D} = (0.99,0.00,0.99)$ for currents entering
through the probe $D$, once again demonstrating almost perfect
pseudospin filtering. The conspicuous absence of current density in
the forward direction indicates that the valley index is well
preserved beyond the path bifurcation point.

The above calculations for current densities and conductances
support the existence of a well defined internal pseudospin degree
of freedom that is well preserved before and after currents are
forced to make turns or split at the channel bifurcation point.
Absence of backscattering and conservation of pseudospin is further
confirmed by calculations carried out in a six-gate geometry in
which a zero line reverses direction as shown in \ref{moreturns}(a).
The pseudospin remains unchanged in which the current propagation
path direction changes from armchair to zigzag and back to armchair
orientations. Similarly results for current-path turn angles of
$60^{\circ}$ and $120^{\circ}$ are shown in \ref{moreturns}(b) and
(c) in which the propagation direction changes from armchair to
armchair.  The absence of an essential role for propagation
direction is further confirmed by the case of a turn angle of
$131^{\circ}$ illustrated in \ref{moreturns}(d). In this case the
current propagation direction after the turn is neither zigzag nor
armchair.


\begin{figure*}
\begin{tabular}{cc}
\includegraphics[width=16cm,angle=0]{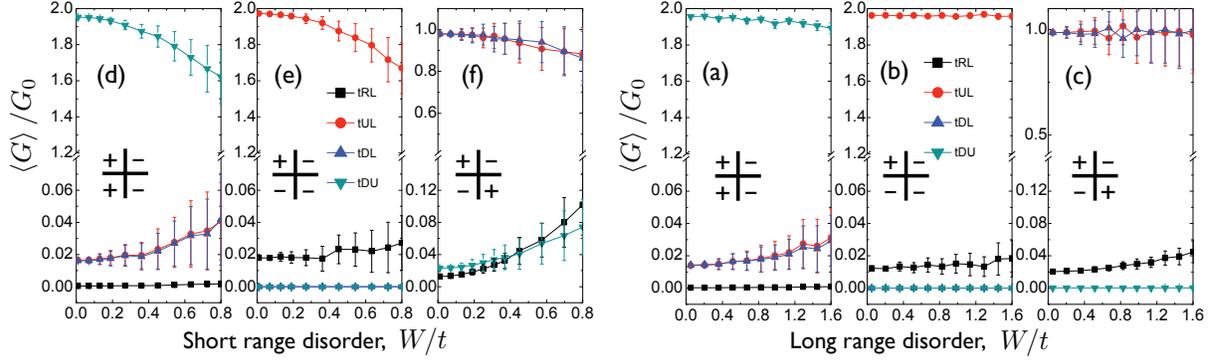}
\end{tabular}
\caption{ (Color Online) Average conductance in units of $G_0=e^2/h$
as a function of external disorder strength for short range onsite
disorder (a), (b), (c) calculated with 200 disorder realizations,
and long range disorder (d), (e), (f) obtained with 100 disorder
realizations. We observe that for disorder strengths comparable to
or smaller than the interlayer bias, the conductances remain very
close to clean-limit values. Deviations of less than $15\%$ are seen
when the disorder energy is twice the interlayer potential
difference and 2/3 of the clean-limit conductivity is retained even
when the disorder strength is four times larger than the interlayer
potential difference. The small leakage of currents to forbidden
channels shown here is largely due to the finite size of our
simulation cell and is expected to be greatly reduced in larger
systems. Note that we have used a different scale to represent those
conductances which have small values in the disorder-free case. }
\label{disorder}
\end{figure*}

Short range disorder sources, for example vacancies, grain
boundaries, or other structural defects, can provide the relatively
large momentum transfers necessary to backscatter kink states along
non-armchair directions. We wish to the assess circumstances under
which disorder can weaken the current-path control based on valley
pseudospin memory discussed in the previous section. In the
following we present numerical results for inter-lead transmission
coefficients in the presence of short-range and long-range disorder
potentials. In order to speed up these calculations we used a
somewhat smaller system size than in the previous section. The
results reported below are for disordered systems with $n_y = 80$
and $n_x=92$ sites. Systems of this size have $N=14970$ atoms in the
central scattering region.

To model short range disorder we add a random contribution to the
on-site potential: $H_{dis} = \sum_i{\omega_i c^\dag_i c_i}$, where
$\omega_i$ is distributed uniformly in the interval [$-W/2,W/2$]
with $W$ characterizing the strength of the disorder. The symmetries
of our model system allow us to focus on the conductances
$G_{L/R/U,D}$ and $G_{LR}$ when currents enter through probe $D$. In
\ref{disorder}, we show the evolution of the average conductance
$\langle G_{pq} \rangle$ as a function of the disorder strength $W$
for the setups (b), (d), (f) shown in \ref{currents}. In panel (a),
we see that $\langle G_{DU} \rangle$ preserves nearly perfect
transmission without fluctuation for disorder strengths $W / t
\in[0,0.2]$, {\em i.e.} when the disorder potential is smaller than
the bulk gap in the constant potential difference regions. When the
disorder strength increases further, $\langle G_{DU} \rangle$ shows
a mild average decrease of around $10 \%$ and larger fluctuations
between disorder realizations. For $W/t = 0.8$, a disorder strength
four times larger than the bulk gap, $\langle G_{DU} \rangle$ can
still reach $75\%$ of its original conductance. The small leakages
of current along the forbidden directions at the bottom of the
figure which appear even in the clean limit reflects insufficient
system size to completely eliminate these tunneling transmission
paths. This quantity is however very small and its further increase
due to scattering introduced by disorder does not surpass 2$\%$ of
the total conductance for the strongest disorder strengths we have
considered.

Similar trends are seen in the other setups examined in panels (b)
and (c), confirming that ballistic transport due to valley
pseudospin memory is extremely robust for disorder strengths smaller
than or comparable to the system gap.

We have also considered disorder models with finite range
correlations of the disorder potential by examining the model $V_{i}
= \sum_{j} \omega_j \, \exp \left( - \left| {\bf r}_j - {\bf r}_i
\right|^2/ 2 \xi^2 \right)$ where we used $\xi = 5a$ in the present
calculations. The sum is carried out over all the neighboring sites
and the local disorder strength $\omega_j$ is bracketed between the
interval [$-\widetilde{W}/2,\widetilde{W}/2$]. The renormalization
of disorder strength by $\widetilde{W} = W / (2 \xi/a)^2$ as a
function of $\xi$ allows a more consistent comparison with the local
disorder model and reflects the relative increase of disorder
strength due to the long range. The above relation was found through
an empirical fitting to data obtained summing the effects of the
Coulomb potential generated by all the surrounding long range
impurity sites with constant strength $W$. The factor $4$ is
slightly smaller than a value of $2\pi$ that we would obtain
analytically from an integration in 2D space. Our numerical results
for this model are illustrated in panels (d)-(f) of \ref{disorder},
where we plot the average conductance $\langle G \rangle$ as a
function of the effective disorder strength $W$. The conductivities
we obtained for long range disorder show an even greater robustness
than the results obtained with short range disorder potentials.

\begin{figure}
\begin{tabular}{cc}
\includegraphics[width=8cm]{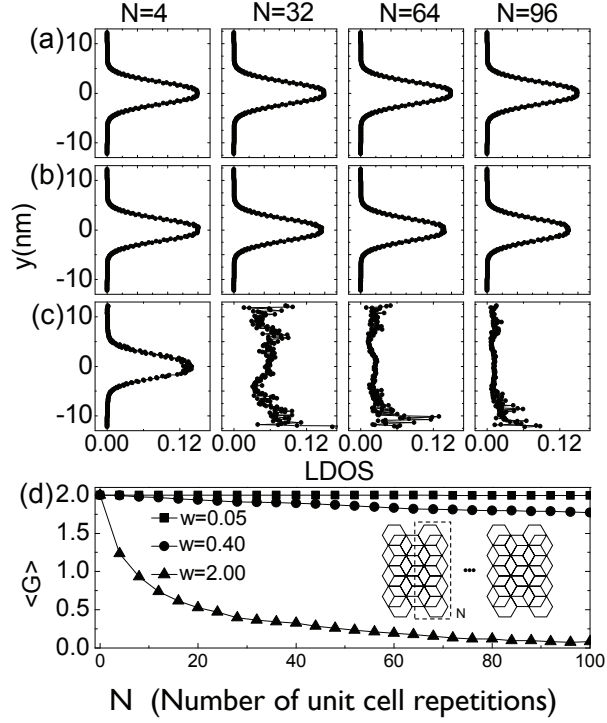}
\end{tabular}
\caption{Transverse local density-of-states profiles for at a series
of positions in the scattering region at a series of points along
the zero-line transmission path for a wide armchair ribbon with a
bulk gap $E_g \sim 0.1 \, t $ and short range disorder strengths of
$W / t = 0.05, 0.4, 2$ in panels (a), (b) and (c) respectively. The
kink states are spread over a width of about $L \sim 10 nm$ and are
not easily backscattered even for disorder strengths comparable to
the band-gap, as shown in panel (d). In the limit of very strong
disorders we can observe the spreading of the kink state wave
functions throughout the whole ribbon and a substantial decrease of
the conductivity $G$ as it propagates. } \label{dosevolution}
\end{figure}

To identify the physics behind the robustness of kink states against
disorder, we have calculated the local density of states as it
propagates along the disordered medium and illustrated our findings
in \ref{dosevolution}. From the calculated local density-of-states
we notice that even for moderately strong disorder comparable with
the band gap, as shown in panel (b), the transverse profile of the
kink states is not modified significantly during transmission.
(There is a decrease in the magnitude of the local density-of-states
upon moving along the zero-line transmission path due to back
scattering.) This robustness of the kink channels to disorder also
explains why the conductance along the forbidden channels
represented in \ref{disorder} hardly increases with disorder
strength. It is only in the limit of very strong disorder as shown
in panel (c), that the 1D character of the kink channels is
completely destroyed, spreading the wave function more homogeneously
across the ribbon. In panel (d) we have plotted the evolution of the
averaged conductivity $\left< G \right>$ as a function of position
obtained by evaluating recursively the resolvent between the first
and the $N$-th unit cell layer in an armchair ribbon. These results
demonstrate that the backscattering rate is fairly small as long as
the kink state retains its original shape.

Further insight into the suppression of backscattering in kink state
transport in the weak disorder limit from an initial kink state $i$
to one of the two final kink states $f$ with opposite velocities can
be gained from the mean free path calculation of an incident state
$i$ using the Fermi's Golden-Rule approximation.
These 1D kink states conducting channels are highlighted by dots in the
corresponding band structures in \ref{Setup}(c) and (d), where we
have designated the right-going states by labels A, B and the left
goers by labels C, D. The transverse eigenvectors of these states
are described by coefficients $c_{i, \tau}$ and $c_{f, \tau}$ where
$i$ = A, B and $f$ = C, D, and $\tau$ labels lattice sites across
the ribbon. With this notation we can measure the degree to which
channels overlap spatially by defining $S_{if} = \sum_{\tau} \left|
c_{i, \tau} \right|^2 \left| c_{f, \tau} \right|^2$.
The Golden rule decay time of a state $\left| k n \right>$  into a set
of final states $\left| k' n' \right>$ can be written as
\begin{eqnarray}
\label{decay}
\tau^{-1}_{k, n}
= \frac{2 \pi}{\hbar} \sum_{k' n'} \left| \left<  k' n' |V| k n \right>
\right|^2 \delta( \varepsilon_{k' n'} - \varepsilon_{k n} ).
\end{eqnarray}
The mean free path is related with the inverse of the decay time
(in units of the unit cell length along the ribbon $\widetilde{a}$) and is given by
 (we refer the reader to the supporting information for more details of the derivation)
\begin{eqnarray}
\frac{{l}_{i}}{\widetilde{a}} = \frac{\upsilon_{i} \tau_{i}}{
\widetilde{a}} = \frac{12 \, \hbar^2}{\widetilde{a}^2 W^2}
\frac{\upsilon_i}{ \sum_{f} S_{if} / \upsilon_{f}} = \frac{C_i}{W^2}
\sim \frac{6 E_g^2 M}{ W^2}. \label{meanfree}
\end{eqnarray}
where we used the relation $\overline{\omega^2} = W^2/12$ for
uniformly distributed disorder strength $\omega$ within an interval
$W$. The rightmost approximate expression was found using $\hbar
\left| \upsilon_{i, f} \right| \sim E_{g} \widetilde{a}$ where $E_g$
is the bulk band gap and $S_{if} \sim 1 / M$. The dimensionless
number $M \sim 290$ for bilayer armchair ribbon kink states
indicates the effective number of transverse lattice sites over
which the transverse density is significant. This number is
typically of the order a few hundred and is expected to become
larger when the kink potential becomes smoother since it is
essentially proportional to the real space width of the kink state.
The coefficients in Eq.~4 evaluated for scattering of the kink
states indicated in \ref{Setup} are $C_A / t^2 = 2.3 \times 10^3 \,
$ and $C_B / t^2 = 1.5  \times 10^3$ for armchair directions. This
argument suggests that the mean free path of a kink channel can be
of the the order of micrometers even when the disorder strength is
comparable with the bulk band gap, and that it will increase
quadratically for weaker disorder.

The reason behind the suppression of backscattering for short range
disorder is the spreading of kink states over several hundred carbon
atoms, which makes it difficult for an isolated scatterer to
globally modify a given kink state and reverse its direction.
Backscattering for smooth long range disorder is even more
efficiently suppressed due to essentially perfect orthogonality
between transverse eigenvectors in different 1D channels. (As
discussed in the supplementary information the pair of AB and CD
kink states when they propagate along the zigzag direction have
symmetric and antisymmetric eigenvector amplitudes manifested in
units of two carbon lattices. The $\pm k$ transverse states for
armchair channels are complex conjugates and mutually orthogonal.)

The mechanism for the suppression of backscattering of kink states
is different from that relevant to the quantum Hall effect in which
counter propagating sates are spatially
separated~\cite{buttikerqhall}, and different from that relevant to
the quantum spin Hall effect in which counter-propagating states do
overlap spatially but are decoupled when time-reversal invariance is
present~\cite{z2kanemele}, but closely related to the mechanism that
supresses backscattering in large-diameter metallic carbon
nanotubes~\cite{nanotube}. This mechanism is unlikely to lead to
perfectly quantized transport in large systems, but is nevertheless
quite effective and can lead to extremely long mean free paths of
the order of hundreds of microns in relatively clean samples.


In summary, an interlayer potential difference in bilayer graphene
can open up a band gap that can be as large as $\sim 0.3$
eV~\cite{gapbilayer}. One interesting feature of the electronic
structure of these electrically tunable semiconductors is the
presence of large Berry curvature peaks of opposite sign sharply
localized near the two Dirac points of the material. The Berry
curvature is associated with the momentum dependence of the Bloch
state sublattice content, {\rm i.e.} with the sublattice pseudospin.
When momentum space is separated into regions centered on the two
valleys these Berry curvatures suggest the presence of nearly
perfectly quantized Hall effects of opposite sign associated with
the two valleys, {\em i.e.} they suggest a valley Hall effect.  In
this paper we have reported on a numerical study bilayer graphene
ribbons in which the inter-layer potential, and hence the Hall
conductances, change sign as a function of position.  Our study
focuses on the chiral edge state channels~\cite{morpurgo} localized
along zero-lines of the inter-layer potential that are associated
with the valley Hall effect.  As long as valley label is preserved
the chiral states provide one-way current transport channels which
can be manipulated by modulating the inter-layer potential profile,
and in particular the paths of its zero-lines.

Our numerical study examines the robustness of these one-way
kink-state transport channels when the transport channel bends
changing the bilayer graphene crystal trajectory. This point
requires the use of a microscopic lattice model and not a continuum
model~\cite{morpurgo} in which valley identity is automatically
retained. Since the kink states are associated mainly with small
well separated regions of momentum space they are expected to
approximate the chiral perfect transmission properties associated
with the quantum Hall effect. Our numerical study shows that
backscattering from a kink state associated with one valley to a
kink state associated with the other valley is small independent of
bend angles in the zero-line path. This property continues to hold
even when portions of the zero-line path follow armchair directions,
along which the two valleys have identical 1D momentum
projections~\cite{chiral}. The small bend resistance values differ
starkly from the case of ordinary semiconductor quantum
wires~\cite{bendres1,bendres2} in which envelopes satisfy
non-relativistic wave equations. We have found essentially zero bend
resistance due to a perfect transmission along zero lines despite
the sharp turns in the current propagation path, unlike in electron
waveguides formed from GaAs-AlGaAs wafers.

We have also demonstrated that one-way conductance through kink
channels is extraordinarily robust against both short-range and
long-range disorder potentials.  We have attributed this behavior to
long kink-channel mean-free-paths in the weak disorder limit. A
Fermi golden-rule analysis, in which the the mean-free-path depends
quadratically on the ratio between the bulk band gap size and
disorder strength, suggests that kink-channel mean-free-path values
from tens to hundreds of microns should be achievable in bilayer
graphene samples of typical mobility. The robustness against even
short range disorder is essentially due to the wide several
nanometer lateral spread of the kink state wave functions which
reduces the effectiveness of lattice scale disorder. This behavior
is analogous to the familiar inverse diameter dependence of the
backscattering probability in metallic carbon
nanotubes~\cite{nanotube}.

Graphene samples obtained through mechanical exfoliation have
relatively few short range disorder defects in the bulk.  We
therefore expect that efficient pseudospin selective beam splitters
and current direction switches can be manufactured based on bilayer
graphene samples accesible by current experimental methods. Such a
possibility would open avenues for exploring new 1D  transport
physics in an experimentally controllable manner.


We acknowledge financial support received from Welch Foundation
grants F-1255 and TBF1473, NRI-SWAN, DOE grant Division of Materials
Sciences and Engineering DE-FG03-02ER45958, and NSF (DMR0906025).
The Computer Center of The University of Hong Kong is gratefully
acknowledged for high-performance computing assistance [supported in
part by a Hong Kong UGC Special Equipment Grant (SEG HKU09)].

$^\dag$zhqiao@physics.utexas.edu;

$^\ddag$jeil@physics.utexas.edu;

$^\S$On leave from University of Texas at Austin.


%
%
%
%
%

\end{document}